\documentclass[graybox]{SNmult}

\usepackage{type1cm}
\usepackage{makeidx}
\usepackage{graphicx}
\usepackage{multicol}
\usepackage[bottom]{footmisc}
\usepackage{newtxtext}
\usepackage[varvw]{newtxmath}

\usepackage{amsmath}
\usepackage{booktabs}
\usepackage{algorithm}
\usepackage[noend]{algpseudocode}

\newcommand{\R}{\mathbb{R}}
\DeclareMathOperator*{\argmin}{arg\,min}
\DeclareMathOperator{\vcol}{col}
\DeclareMathOperator{\subjectto}{\ {s.}{t.}\ }

\newtheorem{ass}{Assumption}
\newtheorem{obs}{Observation}
\newtheorem{rmk}{Remark}
\newtheorem{dfn}{Definition}
\newtheorem{prp}{Proposition}

\graphicspath{{book_chapter_figs/}}
\makeindex

\begin{document}

\title*{Fair Selection of Generalized Nash Equilibria: A~Comparability Perspective}
\titlerunning{Fair Selection of Generalized Nash Equilibria}

\author{Sophie Hall \and Heinrich H. Nax \and Saverio Bolognani}
\authorrunning{Hall et al.}

\institute{%
  Sophie Hall \at ETH Zurich Automatic Control Lab, 8092 Zurich, Switzerland,
  \email{shall@ethz.ch}
\and
  Heinrich H. Nax \at University of Zurich, Behavioral Game Theory, 8050 Zurich, Switzerland,
  \email{hnax@ethz.ch}
\and
  Saverio Bolognani \at ETH Zurich Automatic Control Lab, 8092 Zurich, Switzerland,
  \email{bsaverio@ethz.ch}
}

\maketitle

\abstract*{Generalized Nash equilibrium (GNE) problems model
self-interested agents coupled through their costs and shared
constraints, and arise throughout control applications that allocate a
scarce common resource. Such problems typically admit a continuum of
equilibria, so any implementation must \emph{select} one. This
selection is not neutral: it decides how the cost of the shared
constraint is distributed among the agents, a decision often expected
to be fair. The variational GNE (v-GNE), which charges all agents a
uniform shadow price, is the conventional choice and is frequently
advocated as fair. Through the lens of \emph{comparability} from social
choice theory, we show that the GNE set itself requires no inter-agent
comparability of costs, whereas the v-GNE implicitly assumes cardinal
unit comparability, one of the stronger notions. We then introduce the
fair GNE (f-GNE), which selects an equilibrium by explicitly optimizing
a fairness metric matched to the comparability at hand, and give
constructive routes to compute it for single-constraint and
linear-quadratic games. An electric-vehicle charging game serves as a
running example throughout this self-contained tutorial.
}

\abstract{Generalized Nash equilibrium (GNE) problems model
self-interested agents coupled through their costs and shared
constraints, and arise throughout control applications that allocate a
scarce common resource. Such problems typically admit a continuum of
equilibria, so any implementation must \emph{select} one. This
selection is not neutral: it decides how the cost of the shared
constraint is distributed among the agents, a decision often expected
to be fair. The variational GNE (v-GNE), which charges all agents a
uniform shadow price, is the conventional choice and is frequently
advocated as fair. Through the lens of \emph{comparability} from social
choice theory, we show that the GNE set itself requires no inter-agent
comparability of costs, whereas the v-GNE implicitly assumes cardinal
unit comparability, one of the stronger notions. We then introduce the
fair GNE (f-GNE), which selects an equilibrium by explicitly optimizing
a fairness metric matched to the comparability at hand, and give
constructive routes to compute it for single-constraint and
linear-quadratic games. An electric-vehicle charging game serves as a
running example throughout this self-contained tutorial.
}

\section{Introduction}
\label{sec:introduction}

Game theory provides a natural language to describe the interaction
of multiple self-interested decision makers, or agents, each of whom
minimizes their own cost. In the classical Nash equilibrium problem,
the agents are coupled only through their cost functions: what one
agent pays depends on what the others decide, but each agent is free
to choose any action from their own decision set. In a
\emph{generalized} Nash equilibrium (GNE) problem, the agents are
additionally coupled through shared
constraints~\cite{facchinei2009generalized}: the set of decisions
available to one agent depends on the decisions of the others. This
seemingly technical extension is what makes the model expressive
enough to capture the allocation of a shared but finite resource,
where the joint decision of all agents must respect a common capacity
limit, and whatever one agent consumes is unavailable to the rest.

This modeling feature is the reason why GNE problems appear
throughout control applications that involve resource allocation.
Examples include energy management and peer-to-peer energy
trading~\cite{belgioioso2022distributed, cadre2020peer}, electric
vehicle charging~\cite{ma2016efficient}, demand-side flexibility and
electricity markets~\cite{benenati2023probabilistic}, communication
and network resource allocation~\cite{bassanini2002allocation,
menache2011network}, environmental resources such as shared water
basins~\cite{krawczyk2005coupled, siebert2010groundwater}, and
interactive motion planning for autonomous
vehicles~\cite{lecleach2022algames}. In all of these settings, access
to resources is increasingly controlled automatically by algorithms,
often distributed through large-scale systems, and the individuals
seeking access can be thought of as self-interested agents, each
wanting to fulfill their needs at minimal cost.

A structural feature of generalized games, and the starting point of
this chapter, is that they typically admit not one but infinitely
many equilibria: the GNE solution set is in general a continuum, even
under strong convexity assumptions on the individual
costs~\cite{facchinei2009generalized}. This multiplicity is first of
all a computational nuisance: most algorithms compute \emph{some}
GNE, and which equilibrium they converge to depends on their
initialization~\cite{kaisermayer2021operation,
facchinei2007generalized}, so the properties of the implemented
solution are difficult to characterize \emph{ex ante}. But
multiplicity also poses a more fundamental question, which is the
subject of this chapter: among the many equilibria, which one
\emph{should} be selected and implemented?

Equilibrium selection is not a neutral act. Different GNEs distribute
the burden of the coupling constraint differently: moving from one
equilibrium to another decreases the cost of some agents and
increases the cost of others. Whoever selects the equilibrium (a
system operator, a market platform, a control algorithm) is therefore
deciding, implicitly or explicitly, how costs are shared among the
participants. Moreover, when the resource at stake is a public
infrastructure or a natural resource, there is often an implicit
societal mandate attached to this decision: the selection is expected
to be justifiable as \emph{fair}.

What does it mean for an allocation to be fair? A principled answer
comes from consequentialist ethics through the doctrine of
\emph{welfarism}: moral rightness is determined by the maximization
of aggregate welfare. The framework demands that individual interests
be aggregated impartially, without privileging oneself or particular
groups, and that the best action is the one that maximizes total
welfare~\cite{hall_welfarist_2026, welfarism-arxiv}. Operationally,
this stance reduces the question ``what is fair?'' to the choice of
an aggregation function of the individual costs. Many such functions
have been proposed and used in control: the utilitarian sum, the
Rawlsian maximin, alpha fairness, and inequality indices such as
Atkinson's~\cite{atkinson1970measurement, bertsimas2012efficiency},
Theil's~\cite{theil1967economics}, and
Jain's~\cite{jain1984quantitative}; see~\cite{lan2010axiomatic,
xinyingchen2023guide} for axiomatic and optimization-oriented
overviews and~\cite{cadre2022parametrized, villa2025fair,
neely2008fairness} for control applications.

Within the GNE literature, the selection question has mostly been
settled by convention in favor of the variational GNE (v-GNE): the
refinement of the GNE in which all agents face the same Lagrange
multiplier for the shared constraint, characterized as the solution
of a variational inequality~\cite{facchinei2009generalized,
kulkarni2012variational}. The v-GNE owes its popularity to solid
mathematical foundations: existence and uniqueness hold under broad
assumptions~\cite{rosen1965existence, facchinei2009generalized},
fast semi-decentralized~\cite{paccagnan2018nash, belgioioso2023semi}
and fully-distributed~\cite{yi2019operator, bianchi2022fast}
algorithms are available, and efficiency bounds have been
established~\cite{kulkarni2019efficiency, cadre2020peer}. On top of
these computational merits, the v-GNE is frequently advocated as the
\emph{fair} choice~\cite{dreves2017generalized,
benenati2023probabilistic, bassanini2002allocation}: all agents are
charged the same shadow price for the shared resource, whereas at a
general GNE the shadow prices differ across agents, which is deemed
unfair when individuals are anonymous or indistinguishable to the
planner~\cite{kulkarni2012variational, harker1991generalized}. This
fairness claim is arguably not satisfactory. Equal prices are
meaningful only if the quantities being priced, the agents'
marginal costs, are measured on comparable scales; if the cost
functions are known only up to individual scalings (because they are
self-reported, or elicited in heterogeneous units), the v-GNE
allocation shifts in ways that are hard to justify. A detailed
discussion of the limits of the v-GNE fairness paradigm is given in
the conference paper~\cite{hall_limits_2025}.

This chapter deals with the constructive problem that this critique
uncovers: how to select a GNE based on an explicit, intentional
fairness criterion. The chapter is meant to be a self-contained
tutorial on this fair selection problem, requiring no prior
familiarity with GNE theory.

We approach the problem from a comparability angle, a notion
studied in formal social choice
theory~\cite{roberts1980interpersonal, daspremont2002social}.
Comparability defines to what extent the costs incurred by different
agents are commensurable. By deciding a level of comparability, a
planner decides what information about the individual costs is
relevant to the resource allocation problem. Conversely, a limited
level of comparability can be seen as an invariance condition: it
dictates what information about the individual costs (e.g., scale,
offset) should not affect the resource allocation. See~\cite{welfarism-arxiv}
for a general framework for comparability in control theory.
Adequately modeling comparability is of course particularly crucial
when optimizing for metrics such as fairness, equality, and equity:
because only by defining \textit{a priori} what can be considered
``equal'' can we optimize for any such metric meaningfully.

Through this lens, we show that the solution set of GNEs does not
rely on cardinal comparability of the agents' costs because of its
very nature of being a set of competitive equilibria. However,
v-GNEs require a restrictive notion of comparability, as they
fundamentally and inevitably trade off costs: they are a coordinated
decision, not the result of agents' autonomous decisions. The
fairness of the resulting tradeoff depends on whether the agents'
costs are unit-comparable. This difference between GNEs and v-GNEs
has been pointed out in \cite{facchinei2009generalized}
and \cite{dreves2018how} as a matter of scale invariance, but it had
not been connected to comparability or fairness.

Besides the conceptual analysis, the chapter presents a computational
path to fair GNE selection. The selection problem is formulated as a
bilevel program in which a fairness metric, chosen \emph{a priori}
and consistent with the comparability at hand, is optimized over the
GNE set; we call the resulting solution concept the fair GNE (f-GNE).
For the practically important class of games with a single coupling
constraint, the GNE set admits a tractable parametric description
that makes this bilevel problem solvable with standard tools. For
linear-quadratic games, we go further and show, following the
methodology of~\cite{hall2024explicit}, that the f-GNE can be
precomputed \emph{offline} for all values of the problem parameters
simultaneously (the quantities that vary between instances but are
known at decision time, such as reference setpoints, prices, or the
available charging capacity in the EV example)
via multiparametric programming~\cite{bemporad2002explicit}.
The degree to which this offline precomputation is possible depends
on the fairness metric: for some metrics the result is an explicit
solution map deployable in real time, while for others a residual
online computation remains. This constructive analysis thereby
reveals an interesting interplay between the comparability
assumptions of a fairness metric and the computational tractability
of the associated selection problem.

Throughout, we use an electric vehicle (EV) charging game as a
running example, built up progressively from
Section~\ref{sec:preliminaries} onward.
Section~\ref{sec:preliminaries} introduces GNE and v-GNE definitions
and the ladder of comparability notions.
Section~\ref{sec:GNEcomparability} formally establishes the
comparability requirements of GNE and v-GNE, showing that Nash
equilibria require no comparability while v-GNEs require cardinal
unit comparability.
Section~\ref{sec:fgne} introduces the f-GNE concept, which selects
from the GNE set by optimizing an explicit fairness metric.
Section~\ref{sec:singleconstraint} shows that the selection becomes
tractable for games with a single coupling constraint, by
parametrizing the GNE set through normalized equilibria.
Section~\ref{sec:computation} develops an offline algorithm for
linear-quadratic games and discusses how the comparability
requirements of a fairness metric relate to the computational
tractability of the selection.
Section~\ref{sec:CaseStudy} applies the full framework to the
EV charging game.

\section{Competitive Equilibrium and Comparability}
\label{sec:preliminaries}

\subsection{Generalized Nash Equilibrium Problems}
\label{ssec:solution-concepts-GNE}

Consider a set of self-interested agents $i\in\mathcal{I}:=\{1,\dots,|\mathcal{I}|\}$
each minimizing their cost function $J_i(x_i,x_{-i})$, which depends
on their own decision $x_i \in \R^{n_i}$ and other agents' decisions
$x_{-i}$. We denote by $x$ the vector obtained by stacking all
agents' decisions, i.e., $x = \vcol(\{x_i\}_{i\in \mathcal{I}})\in
\R^n$. Throughout, decisions are deterministic and equilibria are
in pure strategies, the natural setting for continuous games with
strongly convex costs. We will make the following assumption for
the entire chapter.

\begin{ass}[Convex differentiable cost functions]
The cost functions $J_i(x_i,x_{-i})$ are strongly convex and
continuously differentiable in $x_i$, for every fixed $x_{-i}$.
\label{ass:convexJi}
\end{ass}

\noindent Further, each agent's feasible set
\begin{align}
\mathcal{C}_i(x_{-i}):= \{x_i \in \R^{n_i}~|~ g_i(x_i, x_{-i})\leq 0\}
\label{eq:GNEconstraints}
\end{align}
depends on the decisions of other agents through 
$g_i(\cdot): \R^n \to \R^{m_i}$, yielding a set of coupled optimization
problems:
\begin{equation} \label{eq:GNEP1}
\forall i\in \mathcal{I}: \quad \min_{x_i}\, {J_i(x_i, x_{-i})}
\subjectto x_i \in \mathcal{C}_i(x_{-i}).
\end{equation}

\begin{ass}[Convex constraint sets] \label{ass:constraintSets}
The sets $\mathcal{C}_i(x_{-i})$ are closed and convex and the constraint
functions $g_i(\cdot, x_{-i})$ are continuously differentiable.
\end{ass}

Decisions that jointly solve~\eqref{eq:GNEP1} are generalized Nash
equilibria (GNE)~\cite[\S 2]{facchinei2009nash}, as defined next.

\begin{dfn}[Generalized Nash Equilibrium] \label{dfn:GNE}
A joint decision $x^*$ is a GNE of~\eqref{eq:GNEP1} if
$\,\forall i \in \mathcal{I}:$
\begin{align*}
J_i(x_i^*, x_{-i}^*) \leq J_i(x_i, x^*_{-i}), \quad \forall x_i
\in \mathcal{C}_i(x_{-i}^*),
\end{align*}
with the solution set denoted as $\mathcal{S}^{\text{\tiny GNE}}$.
\end{dfn}

Intuitively, a GNE is a joint decision for which no agent
$i \in \mathcal{I}$ can reduce its cost by unilaterally changing its own
decision. 

\begin{svgraybox}
\noindent\textbf{Running Example (EV Charging: Setup).}
Agent $i$ controls a charging action $u_i \geq 0$ to reach a desired
battery level $z_i^{\mathrm{ref}}$ at minimum cost $J_i$. A finite
total supply $\bar{U}$ couples all agents via
$\sum_{j}u_j \leq \bar{U}$. 
The cost $J_i$ has two opposing components: a penalty for
falling short of $z_i^{\mathrm{ref}}$ (incentivizing more charging)
and a congestion price proportional to total demand $\sum_j u_j$
(incentivizing less). Such a demand-dependent price models dynamic energy
pricing schemes in which the unit price of electricity increases
with aggregate demand to incentivize peak shaving, as proposed,
e.g., in~\cite{ma2013decentralized}. Agents thus compete both over
the scarce capacity $\bar{U}$ and over the shared congestion price.
The shortfall penalty is quadratic, so the game is
linear-quadratic; this structure is immaterial for the conceptual
analysis of Sections~\ref{sec:GNEcomparability}
--\ref{sec:singleconstraint}, which holds for general convex costs, but it
will be exploited by the explicit algorithm of
Section~\ref{sec:computation}. The full formulation of this GNE problem appears in Section~\ref{sec:CaseStudy}.
\end{svgraybox}

Under Assumptions~\ref{ass:convexJi}
and~\ref{ass:constraintSets}, finding a GNE is equivalent to solving
the quasi-variational inequality
\begin{align}\label{eq:QVI}
(x-x^*)^\top F(x^*)\geq 0,  \quad \forall \,x  \in \mathcal{C}(x^*)
\end{align}
where $F(x) = \vcol( \{\nabla_{x_i} J_i(x_i,
x_{-i})\}_{i\in \mathcal{I}})$ and
$\mathcal{C}(x) := \prod_{i\in\mathcal{I}} \mathcal{C}_i(x_{-i})$.
Equivalently, $x^*$ is a GNE if a suitable constraint qualification
holds (e.g., the Mangasarian-Fromovitz or Slater's condition) and if
there exist multipliers $\{\lambda_i \in \R^{m_i} \}_{i\in\mathcal{I}}$
such that the following KKT conditions hold:
\begin{align}\label{eq:KKTGNE}
\forall i\in \mathcal{I}:\; \left\{
\begin{array}{ll}
\nabla_{x_i}J_i(x^{*}) + \nabla_{x_i}g_i(x^*)^\top \lambda_i^*=0\\
0\leq \lambda_i^* \perp -g_i(x^*)\geq 0 .
\end{array}
\right.
\end{align}
Here $0\le\lambda_i^*\perp -g_i(x^*)\ge 0$ denotes the complementarity
condition, understood componentwise: for each constraint $k$,
$\lambda_{i,k}^*\ge 0$, $g_{i,k}(x^*)\le 0$, and
$\lambda_{i,k}^*\,g_{i,k}(x^*)=0$, so that a strictly positive
multiplier marks an active (binding) constraint.

In control applications, the majority of works focuses on setups
where the coupling constraints are \textit{jointly
convex}~\cite[Def.\ 3.6]{facchinei2009generalized}, i.e., the
following additional assumption holds.

\begin{ass}[Jointly convex constraints]
There exists a non-empty convex set $\mathcal{K}$ such that
$\mathcal{C}_i(x_{-i}):= \{x_i ~|~ (x_i,x_{-i})\in \mathcal{K} \}$.
\label{ass:jointlyconvexconstraints}
\end{ass}

Notice, in this case, the $g_i(x_i,x_{-i})$ defining
$\mathcal{C}(x_{-i})$ is the same across agents,
$g_i = g, \forall i$, with $g(\cdot): \R^n \to \R^{m}$.
Under Assumptions~\ref{ass:convexJi} and~\ref{ass:jointlyconvexconstraints},
the subclass of \emph{variational} GNEs (v-GNEs) can be defined.

\begin{dfn}[Variational GNE] \label{dfn:vGNE}
Consider the GNE problem~\eqref{eq:GNEP1}, and let
Assumptions~\ref{ass:convexJi}
and~\ref{ass:jointlyconvexconstraints} hold.
The joint decision $x^*$ is a v-GNE if
\begin{align}\label{eq:vGNEsol}
(x - x^{*})^\top F(x^*)\geq 0  \quad \forall x  \in \mathcal{K}
\end{align}
and the solution set is denoted by $\mathcal{S}^{\text{\tiny v-GNE}}$.
\end{dfn}

Equivalently, under appropriate constraint qualifications and
assuming continuous differentiability of the function $g$, $x^*$ is
a v-GNE if there exists a multiplier $\lambda \in \R^{m}$ such that
the following KKT system is satisfied:
\begin{align}\label{eq:KKToriginalvGNE}
\forall i\in \mathcal{I}:\; \left\{
\begin{array}{ll}
\nabla_{x_i}J_i(x^{*}) + \nabla_{x_i}g(x^*)^\top \lambda^{*}=0\\
0\leq \lambda^* \perp -g(x^*)\geq 0
\end{array}
\right.
\end{align}
Note the uniform Lagrange multiplier and the uniform constraint
function $g$ across agents. In general, even in the case of jointly
convex constraints, the set of GNE solutions
$\mathcal{S}^{\text{\tiny GNE}}$ is large, potentially infinite. The
v-GNE is a refinement of the GNE, that is,
$\mathcal{S}^{\text{\tiny v-GNE}}\subseteq
\mathcal{S}^{\text{\tiny GNE}}$~\cite{kulkarni2012variational,
fukushima2011restricted}.

\begin{rmk}
Assumption~\ref{ass:jointlyconvexconstraints} makes the v-GNE
\emph{well-defined}, since all agents share the same constraint
function $g$, but it does not by itself force the multipliers to be
uniform: even under jointly convex constraints, the GNE set contains
equilibria whose KKT conditions~\eqref{eq:KKTGNE} hold with
multipliers $\lambda_i^*$ that differ across agents. Restricting
attention to the solutions with a uniform multiplier is therefore not
a consequence of the agents' individual optimizations; it is an
additional selection criterion imposed by the designer.
Under this criterion, all agents face an identical shadow price for
accessing the shared resource. This has an intuitive fairness
interpretation, but it is meaningful only if agents' costs are
measured on comparable scales. The comparability requirement will be
made precise in Section~\ref{sec:GNEcomparability}.
\end{rmk}

\subsection{Inter-Agent Cost Comparability}
\label{ssec:comparability}

To determine the fairness properties of a joint decision $x$, it is
crucial to compare the cost incurred by individual agents.
In different contexts, it may be that it is meaningful only to compare the absolute costs, to compare the relative improvements, or
to compare costs with respect to a worst-case scenario. What is
comparable in a specific application depends on how difficult it is
to measure certain information, on whether self-reporting can be
trusted, and on what information the social planner considers
relevant (see~\cite{welfarism-arxiv} for further discussion in the
context of control).

This notion of comparability can be expressed by defining a concept
of equivalence between cost functions, i.e., determining what cost functions cannot or should not be distinguished. Equivalently, a specific level of comparability corresponds to a family of transformations $\phi(\cdot)$ of the cost functions that must not affect the joint decision (i.e., for which the decision must be invariant).

\begin{figure}[tb]
\centering
\includegraphics[width=0.8\textwidth]{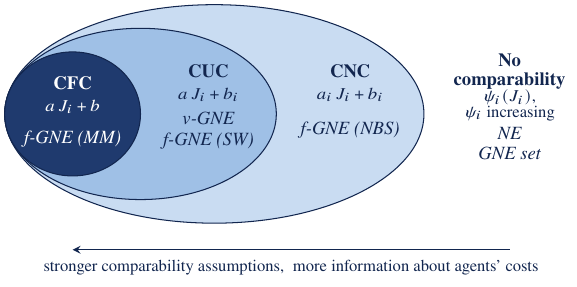}
\Description{Nested families of cost transformations, from
heterogeneous affine transformations (CNC) down to a common affine
transformation (CFC), with arbitrary agent-specific monotone
transformations (no comparability) outside all families, and the
solution concepts that are invariant at each level.}
\caption{Comparability levels read as nested families of cost
transformations, from heterogeneous affine transformations (CNC,
outermost ellipse) down to a common affine transformation (CFC,
innermost); the region outside all ellipses corresponds to
arbitrary agent-specific monotone transformations, i.e., no
comparability. Each solution concept is placed in the
\emph{largest} family of transformations under which it is
invariant: the NE and the GNE set require no comparability
(Observation~\ref{obs:NEordinal}, Remark~\ref{rmk:GNEmonotone}),
the v-GNE requires CUC (Proposition~\ref{prp:CUCofVGNE}), and the
f-GNE (Section~\ref{sec:fgne}) inherits the comparability
requirement of its fairness metric
(Table~\ref{tab:FairnessMetrics}).}
\label{fig:vennGNE}
\end{figure}

In the following, we review a selection of comparability levels
commonly used in social choice
theory~\cite{roberts1980interpersonal}, we report the corresponding
invariance transformation, and we provide a brief interpretation.
Figure~\ref{fig:vennGNE} illustrates the hierarchy of these
comparability levels, together with the solution concepts that will
be associated with each level in the remainder of the chapter.

\medskip

\noindent\textbf{Cardinal Non-Comparability (CNC):}
\begin{align}\label{eq:CNC}
\phi_{\text{\tiny CNC}}(J_i(x)) :=  a_i\,J_i(x) \;+\; b_i,
\quad a_i>0,\; b_i\in\mathbb{R}.
\end{align}
Hence, we cannot compare costs or cost increments across agents, as
the coefficients $a_i$ and $b_i$ are heterogeneous.

\medskip

\noindent\textbf{Cardinal Unit Comparability (CUC):}
\begin{align}\label{eq:CUC}
\phi_{\text{\tiny CUC}} (J_i(x)) :=  a\,J_i(x) \;+\; b_i,
\quad a>0,\; b_i\in\mathbb{R}.
\end{align}
As $a$ is uniform, we can compare cost increments across agents.
However, the absolute cost levels are not comparable.

\medskip

\noindent\textbf{Cardinal Full Comparability (CFC):}
\begin{equation}
\phi_{\text{\tiny CFC}} (J_i(x)) := a\,J_i(x) \;+\; b,
\quad a>0,\; b\in\mathbb{R}.
\end{equation}
A single absolute scale is shared by all agents, so both cost levels
and cost increments are comparable.

\medskip

Moving up this ladder, from CNC to CFC, adds information
requirements and, correspondingly, sensitivity to misreported or
unobservable cost features.

\begin{svgraybox}
\noindent\textbf{Running Example (EV Charging: Comparability Preview).}
Suppose agent~1's cost is scaled by $a_1 = 3$, modeling a vehicle
with a more critical charging need.
Under CNC, this scaling is not considered meaningful information,
and it should not affect which decision is implemented.
Under CUC, the relative scale of the agents' costs is meaningful:
agent~1's cost increments now count three times as much, and the
scaling \emph{should} affect which GNE is selected.
Whether that is desirable depends on whether
$a_1 = 3$ encodes genuine information that has been accurately measured or reported.
\end{svgraybox}

\section{Comparability in Generalized Nash Equilibrium Problems}
\label{sec:GNEcomparability}

We now determine what level of inter-agent comparability the
GNE solution concepts (reviewed in Section~\ref{ssec:solution-concepts-GNE}) implicitly
assume. As a benchmark, consider first the classical Nash
equilibrium problem, in which agents are coupled only through their
costs.

\begin{obs}[NE requires no comparability]
\label{obs:NEordinal}
In a Nash equilibrium problem \emph{without} coupling constraints
($\mathcal{C}_i(x_{-i}) = \mathcal{X}_i$ independent of $x_{-i}$), agent
$i$'s best response depends only on $\nabla_{x_i}J_i$.  Any
strictly increasing transformation $\psi_i: \R \to \R$ applied to
$J_i$ leaves the best response, and therefore the NE solution
set, unchanged.  NE is thus invariant under any agent-specific
monotone transformation of costs, requiring no inter-agent
comparability whatsoever.  Agents in a Nash game are purely
competitive: only the \emph{direction} of each agent's own gradient
matters, not its magnitude relative to others.
The restriction to pure strategies matters here: under mixed
strategies, a nonlinear $\psi_i$ would alter agent $i$'s
preferences over random outcomes, and only positive affine
transformations preserve expected-cost rankings.
\end{obs}

Consider now the generalized game~\eqref{eq:GNEP1} with coupling
constraint sets~\eqref{eq:GNEconstraints} and the problem of finding
$x^{*}\in \mathcal{S}^{\text{\tiny GNE}}$, as stated in
Definition~\ref{dfn:GNE}. Let Assumptions~\ref{ass:convexJi}
and~\ref{ass:constraintSets} hold true. Further, assume that for
every agent $i$, a suitable constraint qualification holds at $x^*$.
The following result relates GNE solutions to the cardinal
comparability of agents' costs.

\begin{prp}[Invariance of GNE under CNC]\label{prp:CNCofGNE}
Let Assumptions~\ref{ass:convexJi} and~\ref{ass:constraintSets}
hold. The solution set $\mathcal{S}^\text{\tiny GNE}$ of the GNE
problem~\eqref{eq:GNEP1} is invariant under
transformations $\phi_{\text{\tiny CNC}} (\cdot)$ defined
in~\eqref{eq:CNC}.
\end{prp}

\begin{proof}
The proof is given in \cite[App.~A]{hall_limits_2025}.
\end{proof}

\begin{rmk}\label{rmk:GNEmonotone}
The mechanism behind Proposition~\ref{prp:CNCofGNE} is visible in
the KKT conditions~\eqref{eq:KKTGNE}: rescaling agent $i$'s cost by
$a_i > 0$ rescales the stationarity condition of agent $i$ only,
and the rescaling can be absorbed by that agent's \emph{own}
multiplier, $\tilde{\lambda}_i^* = a_i \lambda_i^*$, which remains
non-negative. Since each agent has an individual multiplier, the
heterogeneous scalings $a_i$ leave the solution set untouched.
The invariance of the GNE set in fact extends beyond affine maps:
since Definition~\ref{dfn:GNE} involves only within-agent cost
comparisons, the ordinal argument of Observation~\ref{obs:NEordinal}
applies verbatim, and the GNE set is unchanged under any
agent-specific strictly increasing transformation. The restriction
to positive affine transformations in
Proposition~\ref{prp:CNCofGNE} is a matter of class preservation
rather than invariance: a nonlinear monotone $\psi_i$ may destroy
the strong convexity and differentiability required by
Assumption~\ref{ass:convexJi} (e.g., $\psi(s)=\log s$ applied to
$J(x)=x^2+1$), in which case the KKT system~\eqref{eq:KKTGNE} no
longer characterizes the (unchanged) solution set.
\end{rmk}

Proposition~\ref{prp:CNCofGNE} shows that the GNE set is a
fairness-neutral object: it is unchanged when the cost of each
agent is individually rescaled and offset, and therefore it cannot
encode any comparison between the costs incurred by different
agents. This is not the case when we consider the refinement to
v-GNEs, as the following result shows: the v-GNE is not invariant
under heterogeneous rescalings, so the selected solution does
depend on how the agents' costs are scaled against each other.

\begin{prp}[Invariance of v-GNE under CUC]\label{prp:CUCofVGNE}
Let Assumptions~\ref{ass:convexJi}--\ref{ass:jointlyconvexconstraints}
hold. The solution set $\mathcal{S}^{\text{\tiny v-GNE}}$ of the GNE
problem~\eqref{eq:GNEP1} is generally not invariant under
transformations $\phi_{\text{\tiny CNC}} (\cdot)$ defined
in~\eqref{eq:CNC}. $\mathcal{S}^{\text{\tiny v-GNE}}$ is invariant under
$\phi_{\text{\tiny CUC}} (\cdot)$ defined in~\eqref{eq:CUC}.
\end{prp}
\begin{proof}
The proof is given in \cite[App.~B]{hall_limits_2025}.
\end{proof}

The intuition for Proposition~\ref{prp:CUCofVGNE} mirrors the
remark above: in the KKT system~\eqref{eq:KKToriginalvGNE} the
multiplier $\lambda^*$ is shared, so it can absorb a \emph{common}
scaling $a$ (hence invariance under CUC) but not heterogeneous
scalings $a_i$ (hence sensitivity to CNC transformations).

\begin{svgraybox}
\noindent\textbf{Running Example (EV Charging: Illustration of Props.~\ref{prp:CNCofGNE} and~\ref{prp:CUCofVGNE}).}
Consider a two-agent EV game, $\bar{U}=1$, symmetric parameters
(Section~\ref{sec:CaseStudy}).  Both agents receive $u_i = 0.5$ at
the v-GNE. Now scale $J_1 \mapsto 3J_1$.  The GNE \emph{set}
does not change (Prop.~\ref{prp:CNCofGNE}): all allocations
from $u = (0.182, 0.818)$ through $(0.818, 0.182)$ remain equilibria (see Figure~\ref{fig:FairnessMetrics}).
But the v-GNE shifts from $u = (0.5, 0.5)$ to $u \approx (0.659, 0.341)$,
allocating more charge to the agent whose cost was scaled
(Prop.~\ref{prp:CUCofVGNE}).
\end{svgraybox}

The fact that v-GNE, as a solution concept, requires at least
Cardinal Unit Comparability (see Figure~\ref{fig:vennGNE}) is
particularly important when fairness considerations are invoked to
justify it as a refinement of the larger set of GNE solutions. When
doing so, the social planner needs to ensure that the comparability
of cost increments is possible and meaningful. For example, they need
to ensure that (i) the elicitation of cost functions from individuals
is robust against strategic manipulation by the agents; (ii) the
class of admissible cost functions is rich enough to represent the
real cost (more precisely: cost increments) incurred by the agents;
and (iii) that enough information from the agents is collected to
calibrate the relative scaling of each agent's cost function.

We propose an alternative approach: to choose one solution in the GNE set based on an explicit fairness criterion. That is the f-GNE, introduced in the next section. 
Note that by Proposition~\ref{prp:CNCofGNE} the GNE set itself is invariant under CNC transformations, so the search space of the selection is scale-free; whether the \emph{selected} point inherits this invariance depends only on the selection criterion.

\section{Fair Generalized Nash Equilibrium (f-GNE)}
\label{sec:fgne}

As discussed in the Introduction, the set of GNEs is large and
selecting one is necessary in real implementations. A common goal is
to design a solution concept which protects individual self-interest,
ensures strategic stability (thus, a GNE), and \textit{additionally}
optimizes a desired fairness metric consistent with the assumed
inter-agent cost comparability. As pointed out previously, v-GNE, as
a GNE-refinement, does not generally achieve that. Instead, in line
with this observation and inspired by the work
in~\cite{cadre2022parametrized}, we introduce a new approach for
``fair'' equilibrium selection that consists in solving the bilevel
optimization problem
\begin{subequations}\label{eqn:fGNE}
\begin{align}
x^{\text{\tiny f-GNE}} \in \argmin_{x^*}&\quad f(x^*)\\
\subjectto & \quad x^* \in  \mathcal{S}^{\text{\tiny GNE}}
\label{eqn:fGNE_GNEset}
\end{align}
\end{subequations}
where $f$ is a fairness metric that, when minimized, defines the
desired efficiency-equity tradeoff. Selection of GNEs has been
studied in~\cite{benenati2023optimal} from an algorithmic
perspective, in contrast to the fairness perspective considered here.

The standard metrics in Table~\ref{tab:FairnessMetrics}, common in control applications,
represent different points on the comparability ladder introduced
in Section~\ref{ssec:comparability}, as visualized in Figure~\ref{fig:vennGNE}.
All functions are stated as quantities to minimize, consistent with~\eqref{eqn:fGNE}.
Encoding fairness in a scalar aggregation of the individual costs
is itself a design commitment, the welfarist approach of social
choice theory: all information deemed relevant to the selection
enters through the profile of individual costs, and the
aggregation function determines which transformations of those
costs leave the selection invariant. Each metric in
Table~\ref{tab:FairnessMetrics} therefore carries an intrinsic
comparability level, the weakest assumption of
Section~\ref{ssec:comparability} under which its minimizers are
well-defined, reported in the last column. Welfarism is not the
only route to fairness (procedural notions are an alternative),
but it is the one that integrates naturally with the
optimization-based selection~\eqref{eqn:fGNE};
see~\cite{welfarism-arxiv} for a systematic treatment of social
cost functions in control.

\begin{table}[tb]
\centering
\renewcommand{\arraystretch}{2.2}
\begin{tabular}{@{}lcc@{}}
\textbf{Fairness metric} & \textbf{Function} $f(\cdot)$ & \textbf{Comparability}\\
\toprule
Rawlsian maximin (MM)~\cite{rawls1971theory}
  & $\displaystyle \max_{i\in \mathcal{I}} J_i(x)$
  & CFC \\
Social welfare / utilitarianism (SW)
  & $\displaystyle \sum_{i\in \mathcal{I}} J_i(x)$
  & CUC \\
Nash bargaining solution (NBS)~\cite{nash1950bargaining}
  & $\displaystyle -\prod_{i\in \mathcal{I}} \bigl[J_i (\bar x) - J_i(x)\bigr]$
  & CNC \\
Alpha fairness / Atkinson's index (AI)~\cite{atkinson1970measurement,bertsimas2012efficiency} \hspace{3mm}
  & $\displaystyle \sum_{i\in \mathcal{I}} \frac{\left[ J_i(x) \right]^{1 - \alpha}}{1 - \alpha},\; \substack{\alpha \neq 1,\\ \alpha>0}$
  & CFC \\
Jain's index (JI)~\cite{jain1984quantitative}
  & $\dfrac{\bigl(\sum_{i\in \mathcal{I}} J_i(x)\bigr)^2}{|\mathcal{I}| \sum_{i\in \mathcal{I}} (J_i(x))^2}$
  & CFC \\[6pt]
\bottomrule
\end{tabular}
\caption{Fairness metrics and their comparability requirements (see hierarchy in Fig.~\ref{fig:vennGNE} and definitions in Section~\ref{ssec:comparability}).
By minimizing the function $f(\cdot)$, SW minimizes total cost (efficiency); MM optimizes the worst-off
outcome; NBS promotes proportional cost improvements relative to
a benchmark $\bar{x}$; AI interpolates between SW ($\alpha=0$),
NBS ($\alpha=1$) and MM ($\alpha\to\infty$); JI minimizes relative
variations across agents.}
\label{tab:FairnessMetrics}
\end{table}

Interestingly, the v-GNE is a solution of~\eqref{eqn:fGNE} under a
specific problem structure and choice of $f$.

\begin{prp}\label{prp:vGNEequalfGNE}
Let Assumptions~\ref{ass:convexJi}--\ref{ass:jointlyconvexconstraints}
hold and assume the cost functions of agents are fully decoupled,
i.e., $J_i(x)$ is $J_i(x_i)$, and utilitarian fairness is considered
with $f=\sum_{i\in \mathcal{I}} J_i(x)$. Then
$x^{\text{\tiny v-GNE}} = x^{\text{\tiny f-GNE}}$.
\end{prp}
\begin{proof}
The proof is given in \cite[App.~C]{hall_limits_2025}.
\end{proof}

Note that the connection between the v-GNE solution and the social
welfare optimum under decoupled costs has been previously pointed out
in~\cite{cadre2020peer} and~\cite{moret2020loss}.

\begin{svgraybox}
\noindent\textbf{Running Example (EV Charging: Different f-GNEs).}
In the two-agent EV game, the GNE set is a one-dimensional arc on the
boundary $u_1 + u_2 = \bar{U}$ (see Figure~\ref{fig:FairnessMetrics} in Section~\ref{ssec:fairnessmetrics}).  The f-GNE selects a point on this arc.  Under NBS (CNC-compatible), the selection is invariant to cost
scaling by $a_1$.  
Under SW (CUC-required), it shifts with $a_1$. 
Under MM (CFC-required), it moves to equalize individual costs.
\end{svgraybox}

The main difficulty with~\eqref{eqn:fGNE} is that the
constraint~\eqref{eqn:fGNE_GNEset} corresponds to the KKT
system~\eqref{eq:KKTGNE}: when there are multiple coupling
constraints, there exists no closed-form expression of
$\mathcal{S}^{\text{\tiny GNE}}$ and even approximating it numerically is
a challenge~\cite{nabetani2009parametrized}. The next two sections
develop two complementary routes to tractability.
Section~\ref{sec:singleconstraint} treats games with a single
coupling constraint: the GNE set then admits a parametric
characterization that is valid for general convex costs and reduces
the selection to a low-dimensional search.
Section~\ref{sec:computation} treats linear-quadratic games, where
multiple linear coupling constraints can be handled and the
selection can be largely precomputed offline.

\section{Fair Selection under a Single Coupling Constraint}
\label{sec:singleconstraint}

Under Assumption~\ref{ass:jointlyconvexconstraints} (jointly convex
constraints, so $g_i = g$ for every agent), the GNE set admits a
useful parametric characterization via normalized equilibria,
recalled next.

\begin{dfn}[Normalized equilibrium] \cite[Dfn.~3.2]{nabetani2009parametrized}
Let Assumption~\ref{ass:jointlyconvexconstraints} hold. A GNE $x^*$
is called a normalized equilibrium if there exist
$\forall i \in \mathcal{I}: (x^*, \lambda^*_i)$ which solve~\eqref{eq:KKTGNE}
(with $g_i = g$ for all $i$) and $\{r_i>0\}_{i \in \mathcal{I}}$ satisfying
\begin{align*}
r_1 \lambda_1^*= r_2 \lambda^*_2 = \dots = r_{|\mathcal{I}|} \lambda^*_{|\mathcal{I}|}.
\end{align*}
We say $x^*$ is associated with $r = \{r_i\}_{i \in \mathcal{I}}\in
\R^{|\mathcal{I}|}_{>0}$.
\end{dfn}

Under Assumption~\ref{ass:jointlyconvexconstraints}, the v-GNE is the
special case of normalized equilibrium with $r_i = 1\ \forall i$,
consistent with the shared multiplier of~\eqref{eq:KKToriginalvGNE}.

We now specialize to the practically important case of a single
scalar coupling constraint ($g: \R^n \to \R$), which covers the EV
charging game and is required by the following characterization
result.

\begin{prp}\cite[Prp.~3.3]{nabetani2009parametrized}
Let Assumptions~\ref{ass:convexJi}--\ref{ass:jointlyconvexconstraints}
hold, and let the coupling constraint be scalar, i.e.,
$g: \R^n \to \R$ ($m=1$), so that the multipliers $\lambda_i^*$
in~\eqref{eq:KKTGNE} are scalar. If $x^*$ is a GNE satisfying
strict complementarity with some
$\{\lambda_i^*\}_{i \in \mathcal I}$ (i.e.,
$g(x^*) = 0$ implies $\lambda_i^* >0,\ \forall i\in \mathcal{I}$),
then $x^*$ is a normalized equilibrium.
\end{prp}

Consequently, in the case of a single shared constraint, every GNE
satisfying strict complementarity can be represented as a normalized
equilibrium and can therefore be sought by solving the weighted VI
\begin{align}\label{eqn:VI-NNE}
(x - x^{*})^\top F_r(x^*)\geq 0  \quad \forall x:~g(x)  \leq 0
\end{align}
where $F_r(x)= \vcol( \{\nabla_{x_i} r_i J_i(x_i,
x_{-i})\}_{i\in \mathcal{I}})$ is parametrized by the
weights\footnote{Existence and uniqueness results of a solution
of~\eqref{eqn:VI-NNE} usually require (strong) monotonicity of
$F_r(x)$, which may not hold $\forall r\in \R^{|\mathcal{I}|}_{>0}$.} $r$.
GNEs violating strict complementarity need not be represented by any
$r\in\R^{|\mathcal{I}|}_{>0}$. Whether such equilibria can be recovered as limits
of normalized equilibria, or whether restricting the selection to
this family preserves the optimum of a given fairness metric,
requires additional problem-specific analysis.

For problems in which the desired f-GNE lies in this normalized
family,~\eqref{eqn:VI-NNE} replaces the GNE-set
constraint~\eqref{eqn:fGNE_GNEset} with a VI parametrized by $r$.
The following example carries this out for the Nash bargaining
solution.

\subsection*{Example: f-GNE via Nash Bargaining}

In order to reduce the informational requirements for the f-GNE
solution, we choose a fairness metric that is applicable in a setting
of cardinal non-comparability, specifically the Nash bargaining
solution (NBS). Using~\eqref{eqn:fGNE} with the NBS metric ensures
that $x^{\text{\tiny f-GNE}}$ is invariant under CNC transformations
defined in~\eqref{eq:CNC}. Since we consider a single coupling
constraint, the proposition above allows replacing the GNE-set
constraint~\eqref{eqn:fGNE_GNEset} with the parametrized
variational inequality~\eqref{eqn:VI-NNE}, so
that~\eqref{eqn:fGNE} becomes:
\begin{subequations}
\begin{align}
x^{\text{\tiny f-GNE}} \in \; &\argmin_{r, x^*} - \prod_{i\in \mathcal{I}} (J_i(\bar{x}) - J_i(x^*))\\
&\subjectto \; (x - x^{*})^\top F_r(x^*)\geq 0, \; \forall x:
g(x)  \leq 0. \label{eqn:fGNE_NNE_set}
\end{align}
\end{subequations}
This reformulation is computationally tractable because the inner
problem~\eqref{eqn:fGNE_NNE_set} is a standard (for fixed $r$,
strongly monotone) variational inequality, for which efficient
projection-type and operator-splitting solvers are
available~\cite{facchinei2003finite}; the outer search over the
weights $r$ is low-dimensional, since with a single scalar
constraint the weights can be normalized to leave $|\mathcal{I}|-1$ degrees of
freedom. In the two-agent case study of
Section~\ref{ssec:fairnessmetrics} this structure is exploited in
its simplest form: there, the GNE set reduces to a one-dimensional
arc of the constraint boundary, and the NBS metric is minimized by
a scalar line search along the arc.

How to ensure
tractability of~\eqref{eqn:fGNE} for higher-dimensional problems
with multiple coupling constraints remains open in general. The
following section develops a constructive algorithm for the
practically important special case of linear-quadratic games.

\section{Explicit Fair Selection for Linear-Quadratic Games}
\label{sec:computation}

Section~\ref{sec:fgne} formulated the f-GNE selection as a bilevel
problem with the GNE set as a constraint, and
Section~\ref{sec:singleconstraint} made it tractable for a single
coupling constraint, at the price of an online search. In this
section we show that for the special case of linear-quadratic games, the set of GNEs remains tractable even in the presence of multiple constraints.
We consider linear-quadratic games parametrized in a \emph{parameter vector} $p \in \mathcal{P} \subseteq
\R^{n_p}$ that collects the quantities that change between problem
instances but are known at decision time: reference setpoints,
prices, initial conditions, resource capacities (in the EV game,
for example, the charge deficits and the available capacity
$\bar{U}$). We show that the structure
of the GNE set consists in a finite collection of possibly overlapping
critical regions and, within each region, an affine parametrization
of the equilibria.
This structure can be computed \emph{offline} for all parameter
values simultaneously, by combining parametric best-response
analysis with multiparametric quadratic programming
(mpQP)~\cite{bemporad2002explicit, hall2024explicit}. The extent to
which the \emph{selection} of the f-GNE within this structure can
also be precomputed depends on the fairness metric: for social
welfare, a convex and uniquely solvable regional problem yields a
piecewise affine (PWA) candidate law, while for other metrics the
offline computation may yield special-case nonlinear laws or leave a
residual optimization online
(Section~\ref{ssec:LQ-fGNE}).
Unlike the route of Section~\ref{sec:singleconstraint}, this construction does
not rely on normalized equilibria and imposes no restriction on the
number of (linear) coupling constraints.

\begin{rmk}
The algorithm developed here assumes that agents' cost functions are
quadratic and constraints are linear. This structural commitment must be justified independently: if the quadratic model is only an approximation of the true preferences,
the fairness guarantees inherited from the chosen comparability
level are correspondingly approximate. Within the quadratic
framework, the comparability conditions of
Section~\ref{sec:GNEcomparability} translate into conditions on
the relative scaling of the cost matrices $Q_i$.
\end{rmk}

\subsection{Linear-Quadratic GNE Problems}
\label{ssec:LQ-setup}

Consider the specialization of~\eqref{eq:GNEP1} in which each
agent minimizes a quadratic cost, strictly convex in its own decision,
subject to shared
linear inequality constraints parametrized by
$p \in \mathcal{P} \subseteq \R^{n_p}$:
\begin{equation}\label{eq:LQ-agent}
\forall i \in \mathcal{I}:\quad x_i^*(x_{-i},p) = \arg\min_{x_i}\;
\tfrac{1}{2}x^\top Q_i x + (c_i + R_i p)^\top x
\quad\text{s.t.}\quad Ax \leq b + Sp,
\end{equation}
where the own-decision Hessian block satisfies $Q_{i,ii} \succ 0$,
and parameters enter linearly in the cost
(via $R_i$) and in the constraint right-hand sides (via $S$).
This covers initial conditions, reference setpoints, prices, and
resource limits as special cases. The cost matrix $Q_i$ acts on
the full stacked decision $x = \vcol(\{x_j\}_{j\in\mathcal{I}})$,
so agent costs may couple through $x_{-i}$ as in~\eqref{eq:GNEP1}.
Positive definiteness of the full matrix $Q_i$ is not required for a
unique best response.

For fixed $x_{-i}$ and $p$, each agent's subproblem
in~\eqref{eq:LQ-agent} is a QP in which both the decisions of the
other agents and the parameter enter linearly. From the perspective
of agent $i$'s subproblem, the pair $(x_{-i},p)$ therefore plays
the role of a joint parameter vector: as it varies over
$\R^{n-n_i} \times \mathcal{P}$, the optimizer is a continuous PWA
function~\cite{bemporad2002explicit}:
\begin{equation}\label{eq:LQ-BR}
x_i^*(x_{-i}, p) =
H_{i}^{j} x_{-i} + G_i^{j} p + h_i^{j}
\quad \text{if } (x_{-i}, p) \in CR_i^{j},
\quad j = 1,\ldots,N_i,
\end{equation}
where each \emph{critical region} $CR_i^j$ is a polyhedron in
$(x_{-i},p)$-space corresponding to a fixed active set of
inequality constraints, and $N_i \leq 2^{n_A}$, with $n_A$ the
number of inequality constraints in~\eqref{eq:LQ-agent} (the number
of rows of $A$).

\subsection{Equilibrium Critical Regions}
\label{ssec:LQ-CR}

A GNE requires all agents to respond optimally simultaneously.
For each combination $\mathcal{T}_k = (j_1,\ldots,j_{|\mathcal{I}|})$ of
per-agent affine laws, substituting~\eqref{eq:LQ-BR} across
agents yields the \emph{parametric linear system}
\begin{equation}\label{eq:LQ-linearsystem}
M_x^k\, x = M_p^k\, p + M_1^k,
\end{equation}
where the matrices $M_x^k$, $M_p^k$, and $M_1^k$ collect the
best-response coefficients. Depending on its rank and consistency,
this system may admit no solution, a unique solution, or an affine
family of solutions. We denote its solutions by $\hat{x}^k$ and call
them equilibrium candidates associated with $\mathcal C_k$. A
candidate is a GNE only if
$(\hat{x}_{-i}^k,p)\in CR_i^{j_i}$ for every agent $i$, because each
affine best-response law is valid only on its associated critical
region. We distinguish the full-rank and rank-deficient cases below.

\medskip
\noindent\textbf{Full-rank combination.}
When $M_x^k$ is full rank, it produces a unique equilibrium candidate
for the fixed active-set combination $\mathcal C_k$:
\begin{equation}\label{eq:LQ-unique}
\hat{x}^k(p):=(M_x^k)^{-1}(M_p^k p + M_1^k).
\end{equation}
The corresponding \emph{equilibrium critical region} is
\begin{equation}\label{eq:LQ-fullrank-region}
CR_k:=\left\{p\in\mathcal P:
(\hat{x}_{-i}^k(p),p)\in CR_i^{j_i}
\quad\forall i\in\mathcal I\right\}.
\end{equation}
For every $p\in CR_k$, the candidate $\hat{x}^k(p)$ is a GNE.
Computationally, substituting~\eqref{eq:LQ-unique} into the linear
inequalities defining the individual critical regions gives a
polyhedron in $p$-space, whose nonemptiness can be checked by a
feasibility linear program.

\medskip
\noindent\textbf{Rank-deficient combination.}
Let $M_x^k=U_1^k\Sigma_1^k(V_1^k)^\top$ be a compact singular value
decomposition, completed by left and right null-space bases $U_2^k$
and $V_2^k$. If $\operatorname{rank}(M_x^k)=n_M<n_x$, the system is
consistent only at parameters satisfying
\begin{equation}\label{eq:LQ-consistency}
(U_2^k)^\top(M_p^kp+M_1^k)=0.
\end{equation}
At such parameters its candidate solutions form the affine family
\begin{equation}\label{eq:LQ-family}
\hat{x}^k(p,y_2):=x_0^k(p)+V_2^k y_2,
\quad y_2\in\R^{n_x-n_M},
\end{equation}
where
$x_0^k(p)=V_1^k(\Sigma_1^k)^{-1}(U_1^k)^\top(M_p^kp+M_1^k)$ is
one particular solution. Regional membership must be imposed on the
whole candidate family. Define the joint polyhedron
\begin{equation}\label{eq:LQ-jointregion}
\mathcal R_k:=\left\{(p,y_2):\begin{array}{l}
p\in\mathcal P,\quad (U_2^k)^\top(M_p^kp+M_1^k)=0,\\
(\hat{x}_{-i}^k(p,y_2),p)\in CR_i^{j_i}
\quad\forall i\in\mathcal I
\end{array}\right\},
\end{equation}
Every $\hat{x}^k(p,y_2)$ with $(p,y_2)\in\mathcal R_k$ is a GNE
associated with $\mathcal C_k$, and
$CR_k:=\operatorname{proj}_p\mathcal R_k$ is the set of parameters
for which this combination produces at least one GNE. Although
$\mathcal R_k$ is polyhedral, computing $CR_k$ can require polyhedral
projection.

\medskip

There is no guarantee that a given combination $\mathcal C_k$ admits any parameter:
for many combinations the membership conditions are mutually
inconsistent, and the corresponding $CR_k$ is empty; these are simply
discarded during the enumeration. We assume additionally that the
parametrized game admits at least one GNE for every $p\in\mathcal P$,
for example when the joint feasible set is nonempty and compact for
every $p$, so that standard existence conditions
apply~\cite{facchinei2009generalized}. Under this additional
hypothesis, every GNE corresponds to at least one active-set
combination, and the nonempty equilibrium critical regions cover
$\mathcal P$.
Distinct equilibrium critical regions may overlap.
If $p\in CR_k\cap CR_{k'}$, both combinations must be examined. They may yield distinct GNEs, but different combinations may also represent the same equilibrium.

\subsection{The f-GNE as a Multiparametric Program}
\label{ssec:LQ-fGNE}

For a fixed parameter $p$, each full-rank combination with
$p\in CR_k$ contributes one GNE candidate, whereas each rank-deficient
combination may contribute an affine family of GNEs. Within a
rank-deficient combination, the regional selection problem is
\begin{equation}\label{eq:LQ-fGNE}
y_{2,k}^*(p)\in\arg\min_{y_2}\;
f\!\left(\hat{x}^k(p,y_2)\right)
\quad\text{s.t.}\quad (p,y_2) \in \mathcal R_k.
\end{equation}
The affine family~\eqref{eq:LQ-family} spans
only the GNEs that share the active-set combination
$\mathcal{T}_k$.
To compare all applicable combinations, define
\[
v_k(p):=
\begin{cases}
f\bigl(\hat{x}^k(p)\bigr),
& M_x^k \text{ is full rank},\\[1ex]
\displaystyle
\min_{y_2:(p,y_2)\in\mathcal R_k}
f\bigl(\hat{x}^k(p,y_2)\bigr),
& M_x^k \text{ is rank deficient}.
\end{cases}
\]
A global
selector chooses an index in $\arg\min_{k:p\in CR_k}v_k(p)$. Hence
runtime evaluation may require testing several regions and comparing
their candidate values.

Whether~\eqref{eq:LQ-fGNE} admits a PWA explicit solution in $p$
depends on the fairness metric. Substituting~\eqref{eq:LQ-family}
into the metrics of Table~\ref{tab:FairnessMetrics}:
\begin{itemize}
\item \textbf{Social welfare (SW):}
$f^{\text{SW}}(\hat{x}^k(p,y_2))$ is quadratic in $y_2$, and $p$
enters its linear term affinely. Its reduced Hessian is
$
H_k^{\mathrm{SW}}
:=(V_2^k)^\top\left(\sum_iQ_i\right)V_2^k.
$
If $H_k^{\mathrm{SW}}\succ0$, the regional selection
problem~\eqref{eq:LQ-fGNE} is a strictly convex mpQP, and its unique
optimizer $y_{2,k}^{\mathrm{SW}}(p)$ is a continuous PWA function of
$p$ on $CR_k$.
\item \textbf{Maximin (MM):}
Introducing an epigraph variable $t$ gives the regional problem
\[
\min_{y_2,t}\ t
\quad\text{s.t.}\quad
J_i\bigl(\hat{x}^k(p,y_2)\bigr)\leq t
\quad\forall i,\qquad
(p,y_2)\in\mathcal R_k.
\]
Because the epigraph constraints are quadratic in $y_2$, this is a
parametrized quadratically constrained program rather than a standard
mpQP. Even when these constraints are convex, its optimizer is not
generally PWA and no explicit solution is guaranteed. The
regional MM problem may therefore be retained for online solution
unless additional problem-specific structure is available.
\item \textbf{Nash bargaining (NBS):}
Let $\bar{x}$ denote the benchmark allocation used in the NBS metric.
The metric is evaluated over GNEs that weakly improve every agent's
cost relative to this benchmark:
\[
J_i\bigl(\hat{x}^k(p,y_2)\bigr)\leq J_i(\bar{x})
\qquad\forall i.
\]
On this domain,
\[
f^{\mathrm{NBS}}
=-\prod_i\left[
J_i(\bar{x})-J_i\bigl(\hat{x}^k(p,y_2)\bigr)
\right]
\]
is a polynomial of degree at most $2|\mathcal{I}|$ in $y_2$. Its global
minimization is generally nonconvex and cannot be certified from
first-order conditions alone, so a numerical global optimization may
remain online. Nevertheless, the equilibrium regions, particular
solutions $x_0^k(\cdot)$, and null-space bases $V_2^k$ are computed
offline. Each remaining regional problem is posed only in
$y_2\in\R^{n_x-n_M}$, rather than in the full decision vector
$x\in\R^{n_x}$, followed, when necessary, by comparison across
overlapping combinations.
\end{itemize}

\medskip
The complete procedure is summarized in
Algorithm~\ref{alg:fGNE}. For convex, uniquely solvable SW subproblems,
regional candidates can be stored as PWA laws. MM and NBS may retain a
regional optimization online.

\begin{algorithm}
\caption{Explicit f-GNE for linear-quadratic GNE problems}
\label{alg:fGNE}
\begin{algorithmic}[1]
\State \textbf{Input:} LQ-GNE~\eqref{eq:LQ-agent}, parameter set
$\mathcal{P}$, and fairness metric $f$.
\State \textbf{Output:} Regional explicit laws and retained online
problems.
\Statex
\For{each agent $i \in \mathcal{I}$}
  \State Solve mpQP~\eqref{eq:LQ-agent}.
  \State Store the PWA response~\eqref{eq:LQ-BR} and regions
  $\{CR_i^j\}$.
\EndFor
\For{each combination $\mathcal{T}_k = (j_1,\ldots,j_{|\mathcal{I}|})$}
  \State Form parametric linear system~\eqref{eq:LQ-linearsystem}.
  \If{$M_x^k$ full rank}
    \State Compute the affine candidate law~\eqref{eq:LQ-unique}.
    \State Compute the region $CR_k$.
  \Else
    \State Compute $U_2^k,V_2^k$ via SVD.
    \State Construct $\mathcal R_k$ as in~\eqref{eq:LQ-jointregion}
    and project it onto the $p$-space to obtain $CR_k$.
    \If{$f=f^{\mathrm{SW}}$ and $H_k^{\mathrm{SW}}\succ0$}
      \State Compute the PWA law $y_{2,k}^{\mathrm{SW}}(p)$.
    \Else
      \State Leave the regional selection problem for online solution.
    \EndIf
  \EndIf
  \If{the regional minimizer is available explicitly}
    \State Store $CR_k$, the regional minimizer, and $v_k(p)$.
  \Else
    \State Store $CR_k$ and the regional problem for online solution.
  \EndIf
\EndFor
\State \Return each $CR_k$ with its explicit solution laws or retained
  regional problem.
\end{algorithmic}
\end{algorithm}

At runtime, for the observed parameter $p$, one identifies all regions
$CR_k$ containing $p$, evaluates the stored candidate laws or solves
the retained regional problems, and selects a candidate with the
smallest value of $f$.

\section{Case Study: Electric Vehicle Charging Game}
\label{sec:CaseStudy}

Consider an electric vehicle charging game, a simplified version
of~\cite{ma2016efficient, ma2013decentralized, liu2018optimal}, as
played by a finite number of agents $i\in \mathcal{I}$. Every agent
arrives with an initial charge $z^{\text{init}}_i$, and charges its
vehicle by $u_i$, aiming for a desired final charge $z^{\text{ref}}_i$.
The aggregate charge is constrained by an available capacity $\bar U$,
which may be insufficient to charge every agent to its desired level.
This leads to the coupled optimal control problems
\begin{align}\renewcommand{\arraystretch}{1.5}
\forall i \in \mathcal{I} \left\{\begin{array}{rl}
\displaystyle \min_{u_i, z_i} & q_i\,\|z^{\text{ref}}_i- z_i\|^2
  + u_i\, \sigma(u)\\
\subjectto & z_i = A_i z^{\text{init}}_i + B_i u_i,\\
& 0\leq u_i,\quad \sum_{j\in \mathcal{I}} u_j \leq \bar U,
\end{array}\right.
\end{align}
where $A_i\in [0, 1]$ are the battery leakage rates, and
$B_i \in [0, 1]$ is the charging efficiency. Similarly
to~\cite{ma2013decentralized}, we model the congestion price
$\sigma$ as an affine function of the total
demand\footnote{A congestion penalty or price is commonly used in
multi-step horizon problems to ensure peak shaving or valley
filling. In some formulations the tariff $\rho^1$ is heterogeneous
across agents to model different priority levels; here we keep it
uniform, consistent with a price set by the energy market.}, i.e.,
\begin{align}
\sigma(u) = \rho^1\sum_{j\in \mathcal{I}} u_j + \rho^0,
\end{align}
where $\rho^0$ is a base price and $\rho^1$ a congestion tariff.
Both are set by the energy pricing mechanism and are therefore
identical for all agents; the individual preference of each agent,
namely how much they value reaching their desired charge
$z^{\text{ref}}_i$ relative to the incurred energy cost, is
entirely captured by the weight $q_i>0$.

Applying the KKT conditions of Definition~\ref{dfn:vGNE}
(eq.~\eqref{eq:KKToriginalvGNE}) to the EV cost yields the
pseudo-gradient $F(u) = \Phi u + \varphi$ with
\begin{align*} F(u) \! = \!\!
\left[\begin{smallmatrix}
2(q_1 (B_1)^2 + \rho^1) & \rho^1 & \hdots\\
 & \ddots & \\
\rho^1 & \hdots &   2( q_{|\mathcal{I}|} B_{|\mathcal{I}|}^2 + \rho^1)
\end{smallmatrix}\right] \! u \! + \!\!\left[\begin{smallmatrix}
-2 q_1 \Delta z_1 B_1 + \rho^0\\
\vdots\\
-2 q_{|\mathcal{I}|} \Delta z_{|\mathcal{I}|} B_{|\mathcal{I}|} + \rho^0
\end{smallmatrix}\right]\!,
\end{align*}
where $\Delta z_i = z^{\text{ref}}_i - A_i z^{\text{init}}_i$ is the
effective charge deficit.  The v-GNE satisfies
$F(u) + \mathcal{N}_{\R_+^{|\mathcal{I}|}}(u) + \lambda \mathbf{1} = 0$,
$0 \leq \lambda \perp -(\mathbf{1}^\top u - \bar{U}) \geq 0$.
The parameter vector $p = (\Delta z, \rho^0, \bar{U})$ collects all
quantities that enter this system: under CUC the absolute values of
the weights $q_i$ are meaningful information; under CNC each
agent's cost is defined only up to an individual positive
rescaling, and no meaning can be attached to the relative
magnitude of the $q_i$ across agents.

\subsection{Transformations}
\label{ssec:EVtransformations}

\begin{figure}[t]
\centering
\includegraphics[width=0.95\textwidth]{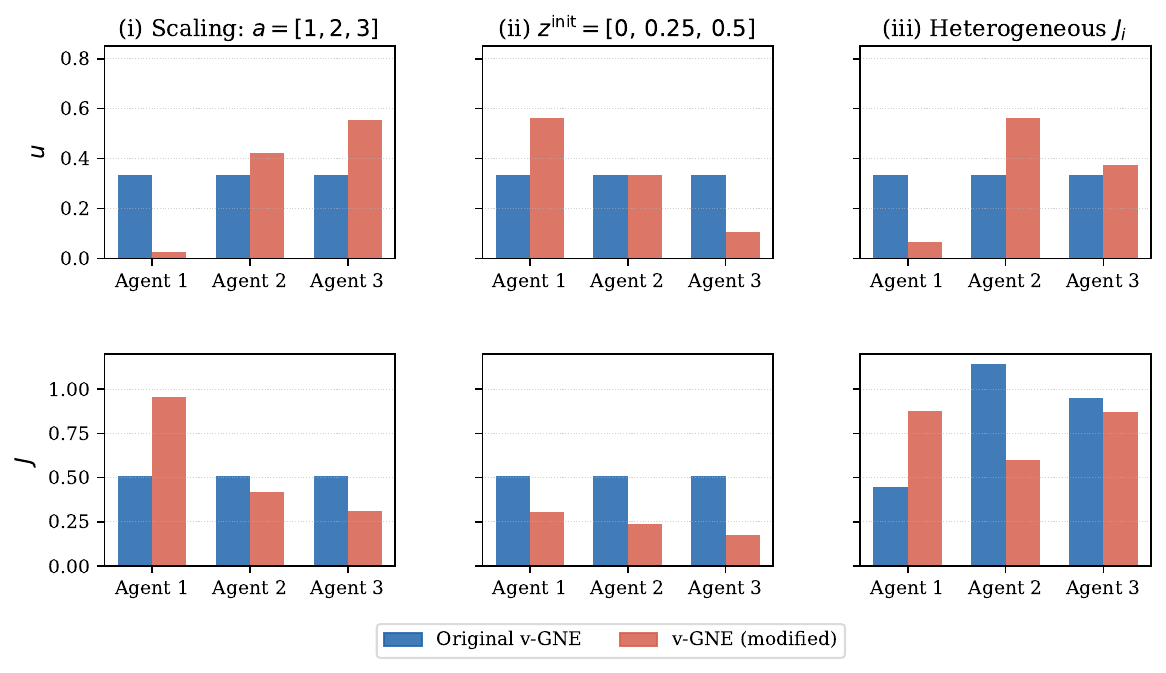}
\Description{Comparison of the v-GNE solution under transformations
of cost scaling, initial condition and cost function structure.}
\caption{Comparison of the v-GNE solution and how it changes under
transformations of cost scaling, initial condition and cost function
structure.}
\label{fig:TransformationsvGNE}
\end{figure}

We now demonstrate Proposition~\ref{prp:CUCofVGNE} concretely.
Three transformations of the EV cost functions are applied; each
changes the v-GNE, illustrating that the designer must decide
whether the information driving the shift is physically meaningful
and should affect the allocation. For each example, we recompute
the v-GNE and report the resulting costs and allocations in
Figure~\ref{fig:TransformationsvGNE}.

\medskip

\noindent\textbf{(i) Scaling of cost functions:}
$\tilde{J}_i(u) = a_i J_i(u)$ with scaling vector
$a = [1\,,\,2\,,\,3]^\top$. Heterogeneous scaling could represent
the case in which the agents' utilities depend on the additional
kilometer range they get from charging their vehicle, and different
vehicles have different fuel economy. Would we really deem it
desirable that our solution concept allocates more charge to less
efficient vehicles?

\medskip

\noindent\textbf{(ii) Change in the initial charge level} of agents
from $z^{\text{init}} = [0\,,\,0\,,\,0]^\top$ to
$\tilde{z}^{\text{init}} = [0\,,\,0.25\,,\,0.5]^\top$.
The share of charge for agent 1 who arrives with an empty battery
increases significantly. However, is it really desirable that agents
with lower initial charge are advantaged under the v-GNE?

\medskip

\noindent\textbf{(iii) Transformed cost:} the cost function for
reaching the desired charge level $z^{\text{ref}}$ is modeled as a
quadratic function $J_{i} = \|z^{\text{ref}}_i- z_i\|^2$. What if
the true utility is heterogeneous across agents? For instance, one
agent needs to reach home but any additional charge will result in
small benefit, represented by a logarithm. Another agent has to
drive a long distance urgently. Consider the following cost
functions:
\begin{align*}
&J_{1}(z_1) = (z_1^{\text{ref}}-z_1)^2,\\
&J_{2}(z_2) = \log\!\left(\frac{\kappa}{z_2}\right) + 0.1 (z_2^{\text{ref}}-z_2)^2,\\
&J_{3}(z_3) = e^{z_3^{\text{ref}}-z_3} - 1.
\end{align*}
Is it desirable that the declared need in terms of utility has such a
strong effect on the allocation?

The examples show that the v-GNE depends both on cardinal cost
representations and on the preference information encoded in the
model. Whether these features should influence the allocation is a
design choice.

\subsection{Fairness Metrics}
\label{ssec:fairnessmetrics}

We now showcase the underlying comparability assumptions of a variety
of fairness metrics $f(\cdot)$ and their susceptibility to changes in
the scale of the cost functions. In order to do so, we compare the
solution of the f-GNE in~\eqref{eqn:fGNE} under three different
\emph{a priori} valid choices of fairness metrics $f(\cdot)$: (i) social welfare (SW),
(ii) maximin (MM), and (iii) Nash bargaining solution (NBS), as introduced in Section~\ref{sec:fgne}.

\begin{figure}
\centering
\includegraphics[width=\textwidth]{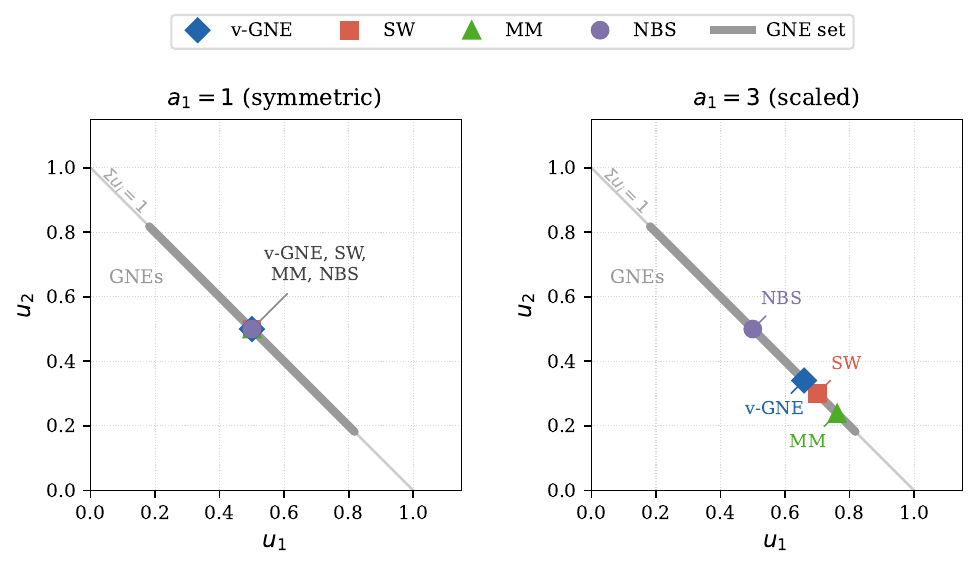}
\Description{Various fairness metrics, the v-GNE, and how their
solution changes with scaling.}
\caption{Various fairness metrics, the v-GNE, and how their solution
changes with scaling $a_1 = \{1,3\}$, left to right. Note that on
the left all solutions overlap.}
\label{fig:FairnessMetrics}
\end{figure}

Specifically, we consider a situation with two agents in which one
unit of charge is available, i.e.\ $\bar U=1$, and initially both
agents are equivalent and thus receive the same amount of charge
$u_i = 0.5,\ i\in \{1,2\}$ as shown in
Figure~\ref{fig:FairnessMetrics}. Then, in a second setting, we scale
the cost function of the first agent by a factor of 3 such that
$\tilde{J}_1(u) = a_1 J_1(u) = 3\, J_1(u)$. Based on
Figure~\ref{fig:FairnessMetrics}, several observations are noteworthy:

\begin{itemize}
\item As stated in Proposition~\ref{prp:CNCofGNE}, the set of GNEs
is invariant under CNC, and thus under scaling of cost functions.
\item In the scaled situation, the maximin fairness criterion selects
the GNE that allocates the most resources to agent 1 which is worse
off under the scaled cost.
\item Different fairness criteria produce different allocations, with
the exception of the Nash bargaining solution, as it is invariant
under CNC transformations.
\end{itemize}

\subsection{Explicit f-GNE for the EV Charging Game}
\label{ssec:EVexplicit}

Algorithm~\ref{alg:fGNE} applies directly to the EV game:
the costs are quadratic, the constraint $\mathbf{1}^\top u \leq \bar{U}$
is linear, and non-negativity $u_i \geq 0$ provides individual bounds.

With parameter vector $p = (\Delta z, \rho^0, \bar{U})$, when the
resource constraint is active the matrix $M_x^k$ has a
$|\mathcal{I}|-1$ dimensional null space tangent to the active resource
hyperplane, expressing that charge can be shifted between agents while
preserving its total. Thus $(V_2^k)^\top\mathbf{1}=0$ and
$y_2\in\R^{|\mathcal{I}|-1}$ in~\eqref{eq:LQ-family}. In the two-agent numerical
experiment below this reduces to a scalar parameter, with tangent
direction $(1,-1)/\sqrt{2}$. We therefore have the following decision
rules under the different fairness criteria:
\begin{itemize}
\item \textbf{SW:} minimizing the aggregate quadratic cost over
$y_2$ subject to the critical-region bounds is an mpQP,
yielding a piecewise-affine (PWA) allocation law
$u^*(p) = u_0^k(p) + V_2^k y_{2,k}^{\mathrm{SW}}(p)$ under the convexity
and uniqueness conditions stated in Section~\ref{ssec:LQ-fGNE}.
\item \textbf{MM:} the minimax conditions (equalization of the worst-off agents' costs,
or activity of a region bound) can be solved in closed form for the
two-agent regimes considered here, yielding a piecewise nonlinear law.
This special-case construction does not imply an explicit MM law in general.
\item \textbf{NBS:} the objective is polynomial in $y_2$; no
general closed-form law exists, and in the two-agent experiment a
scalar numerical global minimization over the GNE arc is performed
online for each parameter instance.
\end{itemize}

\begin{figure}[tb]
  \centering
  \includegraphics[width=0.85\textwidth]{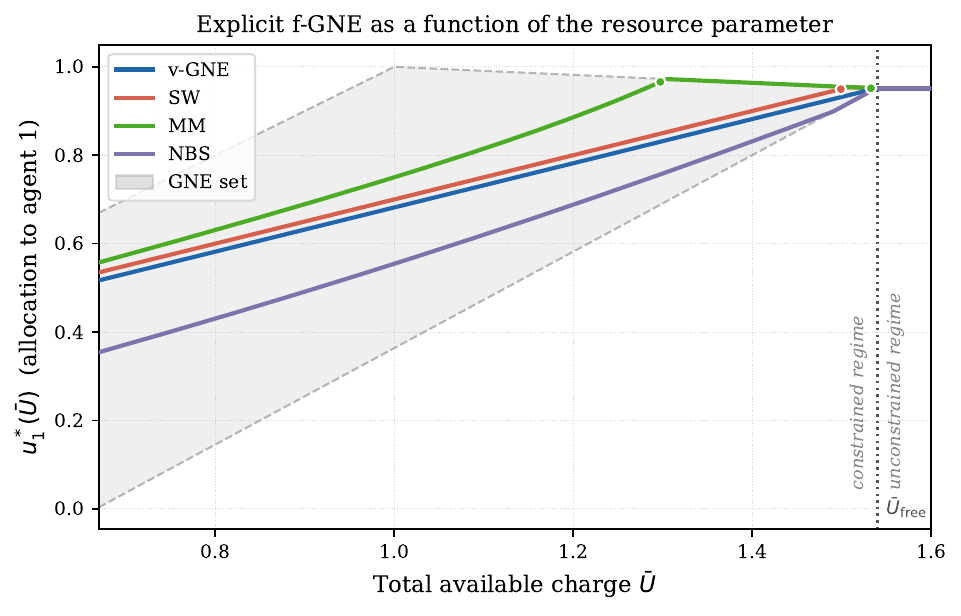}
  \Description{Line chart of agent~1's allocation versus total
    available charge. At each capacity, the GNE set permits an
    interval of allocations for agent~1, within which the v-GNE,
    social-welfare, maximin, and Nash-bargaining criteria select
    particular allocations. In the constrained regime, maximin
    initially gives agent~1 the largest allocation, Nash bargaining
    the smallest, and the v-GNE and social-welfare trajectories lie
    close together. All four trajectories converge near the
    unconstrained-capacity threshold and coincide thereafter.}
  \caption{f-GNE allocation $u_1^*$ for agent~1 as a function of
    available capacity~$\bar{U}$ (two-agent asymmetric EV game,
    $\Delta z_1=1.2$, $\Delta z_2=0.8$).
    The NBS disagreement benchmark is the no-charge allocation
    $(u_1,u_2)=(0,0)$.
    Gray band: range of agent~1's allocations over the GNE set.
    Dotted vertical line: unconstrained threshold $\bar{U}_{\mathrm{free}}$.
    The v-GNE curve is affine and the SW curve piecewise affine in
    $\bar{U}$; the MM and NBS curves are piecewise nonlinear.}
  \label{fig:explicitfGNE}
\end{figure}

Figure~\ref{fig:explicitfGNE} illustrates the difference between these f-GNEs by sweeping
the available capacity $\bar{U}$ for an asymmetric two-agent game
($\Delta z_1 = 1.2$, $\Delta z_2 = 0.8$).
For the NBS selection in this simulation, the disagreement benchmark
is the no-charge allocation $\bar{x}=(u_1,u_2)=(0,0)$, so that each
agent's bargaining gain is measured relative to its cost when neither
agent charges.
For $\bar{U}$ below the unconstrained threshold
$\bar{U}_{\mathrm{free}} \approx 1.54$, the coupling constraint is
active and the GNE set is a non-trivial arc (gray band); above this
threshold the unique Nash equilibrium coincides across all metrics.
Several features of the figure deserve comment.

The \emph{upper boundary of the GNE set} equals $\bar{U}$ for
$\bar{U} \leq 1$ (agent 1 can receive the entire supply) but
decreases afterwards. The cap is agent 1's own rationality: at a
GNE, agent 1 must not prefer to unilaterally reduce its charge, and
beyond the cap the congestion price, which grows with total demand
$\bar{U}$, outweighs the shrinking shortfall benefit.

The \emph{v-GNE curve} is exactly affine in $\bar{U}$ (slope
$0.5$) over the whole constrained regime. The \emph{SW curve} is
piecewise affine, as predicted by the mpQP analysis: it runs
parallel to the v-GNE (slope $0.5$, constant offset) while the
welfare optimum is interior to the GNE set, and it exhibits a kink
shortly before $\bar{U}_{\mathrm{free}}$, where the optimum reaches
the upper boundary of the GNE set and follows it. The constant
offset between SW and v-GNE, despite both being CUC-based, is caused by the fact that the EV costs are coupled through the congestion price, so the
decoupled-cost hypothesis of Proposition~\ref{prp:vGNEequalfGNE}
fails, and the SW selection (which internalizes the externality)
differs from the v-GNE (which equalizes private marginal costs).

The \emph{MM curve} is not affine: in its first regime it
equalizes the two agents' costs, which yields a nonlinear
(rational) expression in $\bar{U}$, and at $\bar{U} \approx 1.3$
it kinks onto the upper boundary of the GNE set.
The \emph{NBS curve} is smooth and nonlinear throughout the
constrained regime for the no-charge disagreement benchmark used here.
All four trajectories converge at $\bar{U}_{\mathrm{free}}$,
confirming that a unique Nash equilibrium exists once the resource
constraint becomes slack.

\section{Discussion and Conclusion}
\label{sec:conclusions}

The v-GNE retains the desirable property that agents face the same ``shadow
price'' for accessing a shared resource. In such a setting, it can be
interpreted as a coordination mechanism analogous to the emergence of
a single clearing price from bids and asks in a competitive market.

However, this interpretation rests on an assumption that the v-GNE
makes implicitly: the agents' costs must be unit comparable. Whether
this assumption is defensible is not a mathematical question. Some
comparability levels require cost information that may be difficult
to quantify, expensive to measure, private, or susceptible to
misreporting. A designer may also wish to prevent particular modeled
features from influencing the allocation. The examples of
Section~\ref{ssec:EVtransformations} illustrate how both cardinal cost
representations and preference information, such as initial charge
and utility shape, can affect the selected outcome.

The conclusion is not that fairness should always be pursued under
the weakest possible comparability requirement. Maximin, for example,
requires full comparability and provides a compelling criterion when
the relevant information is meaningful and trustworthy. The Nash
bargaining metric, at the other extreme, is invariant to individual
cost rescalings and is suitable when less comparability can be
defended. The fairness criterion should therefore be chosen
intentionally and made explicit, in accordance with the information
that is credible in the application. It may differ from the implicit
selection embodied by the v-GNE.

Computationally, games with one coupling constraint admit a
parametrization by normalized equilibria that reduces fair selection
to a lower-dimensional search. For linear-quadratic games with multiple
linear constraints, parametric best-response analysis and active-set
enumeration provide regional parametrizations of the GNE set.
Strictly convex SW selection yields explicit PWA regional laws,
whereas MM and NBS may retain regional optimization online.
Improving the scalability of this construction and extending it to
nonquadratic games with multiple coupling constraints remain open
directions.

Altogether, fair GNE selection remains a rich problem at the
intersection of comparability, invariance, equilibrium selection, and
computation.

\begin{acknowledgement}
This work was supported by the Swiss National Science Foundation
under the NCCR Automation (grant 51NF40\_225155) and under the
Eccellenza Grant ``Markets and Norms'' (H.H.N.).
\end{acknowledgement}

\ethics{Competing Interests}{The authors have no conflicts of
interest to declare that are relevant to the content of this chapter.}

\bibliographystyle{splncs04}
\bibliography{chapter_fairgne}

\end{document}